\documentclass[11pt]{article}
\usepackage{amsmath, amssymb, amsfonts, amsthm}

\usepackage[T1]{fontenc}

\usepackage{mathtools}
\usepackage{tikz}

\usetikzlibrary{calc}
\usetikzlibrary{intersections} 
\usetikzlibrary{patterns,shapes,arrows.meta}

\usepackage{pgfplots}
\usepackage{pgfplotstable}
\usepackage{pgfmath}
\pgfplotsset{compat=1.18}  

\usetikzlibrary{pgfplots.fillbetween}
\usetikzlibrary{plotmarks}

\usepackage{pgfplots}
\usepackage{booktabs}
\usepackage{xcolor}
\usepackage{caption}
\usepackage{setspace}
\usepackage{hyperref}
\usepackage{url}
\usepackage{fancyhdr}
\usepackage{geometry}
\usepackage{array}
\usepackage{enumitem}

\usepackage{erewhon}
\usepackage{euler}

\usepackage[numbers]{natbib}

\usepackage{xcolor}

\definecolor{electricpurple}{rgb}{0.75, 0.0, 1.0}
\definecolor{brightcyan}{rgb}{0.0, 1.0, 1.0}
\definecolor{hotmagenta}{rgb}{1.0, 0.11, 0.81}

\hypersetup{
    colorlinks=true,
    linkcolor=electricpurple,    
    urlcolor=brightcyan,     
    citecolor=electricpurple 
}

\DeclareMathOperator{\E}{\mathbb{E}}
\DeclareMathOperator{\Var}{\mathrm{Var}}

  \makeatletter
\def\@fnsymbol#1{\ensuremath{\ifcase#1\or \ddagger\or \ddagger\or
   \mathsection\or \mathparagraph\or \|\or **\or \dagger\dagger
   \or \ddagger\ddagger \else\@ctrerr\fi}}
    \makeatother

\title{\textbf{On Bessel’s Correction:} \\ \Large Unbiased Sample Variance, the \textit{Bariance}, and a Novel Runtime-Optimized Estimator}
\author{Felix Reichel\thanks{Department of Economics, Johannes Kepler University Linz, 4040 Linz, Austria. Corresponding author. Email: \url{mailto:kontakt@felixreichel.com}, \url{mailto:felix.reichel@jku.at}. 
This research has benefited from 4 open access commentators on \url{https://www.qeios.com/read/3GJGNA}. A replication package for reported empirical runtimes has been published on \texttt{GitHub} under: \url{https://github.com/felix-reichel/BarianceVariance_Reproduction_Repo_Robustness_Supplements}.
}}

\date{June 14, 2024; Last Revised May, 2025}

\begin{document}

\setstretch{1.0}


\maketitle

\begin{abstract}
\noindent Bessel’s correction adjusts the denominator in the sample variance formula from $n$ to $n - 1$ to ensure an unbiased estimator of the population variance. This paper provides rigorous algebraic derivations, geometric interpretations, and visualizations to reinforce the necessity of this correction. It further introduces the concept of \textit{Bariance}, an alternative dispersion measure based on pairwise squared differences that avoids reliance on the arithmetic mean.
\noindent Building on this, we address practical concerns raised in Rosenthal’s article~\cite{rosenthal2015}, which advocates for $n$-based estimates from a mean squared error (MSE) perspective—particularly in pedagogical contexts and specific applied settings. Finally, the empirical component of this work, based on simulation studies, demonstrates that estimating the population variance via an algebraically optimized \textit{Bariance} approach can yield a computational advantage. Specifically, the unbiased \textit{Bariance} estimator can be computed in linear time, resulting in shorter run-times while preserving statistical validity.

\end{abstract}

\newpage
\tableofcontents
\newpage

\begin{center}
\vspace{.25em}

\itshape
``This could be the end of it all,\\
released from the pain.\\
So self-indulgent and so insincere,\\
I’ll never bow to your lies again.''

\vspace{.25em}

--- Megadeth, *Tornado of Souls*
\end{center}

\vspace{.75em}

\noindent \textbf{The following notation is used throughout the paper:}\\
\noindent
\( X_1, X_2, \dots, X_n \in \mathbb{R} \) \dotfill Independent and identically distributed sample from a population. \\[0.5em]
\( \mu = \E[X_i] \) \dotfill Population mean. \\[0.5em]
\( \sigma^2 = \Var(X_i) \) \dotfill Population variance. \\[0.5em]
\( \bar{X} = \frac{1}{n} \sum_{i=1}^n X_i \) \dotfill Sample mean. \\[0.5em]
\( S^2 = \frac{1}{n} \sum_{i=1}^n (X_i - \bar{X})^2 \) \dotfill Biased sample variance estimator (denominator \(n\)). \\[0.5em]
\( \hat{S}^2 = \frac{1}{n-1} \sum_{i=1}^n (X_i - \bar{X})^2 \) \dotfill Unbiased sample variance estimator (Bessel-corrected). \\[0.5em]
\(\text{\textit{Bariance}} = \frac{1}{n(n-1)} \sum_{i \neq j} (X_i - X_j)^2\) \dotfill Pairwise variance estimator using all ordered pairs (each unordered pair counted twice). \\[0.5em]
\(\text{\textit{Bariance}}_{\text{opt}} = \frac{2n}{n(n-1)} \sum_{i=1}^n X_i^2 - \frac{2}{n(n-1)} \left( \sum_{i=1}^n X_i \right)^2\) \dotfill Optimized scalar formula for \textit{Bariance}. \\[0.5em]
\( \sum_{i < j} (X_i - X_j)^2 \) \dotfill Sum over all unordered pairwise squared differences (each pair counted once). \newline Note: \(\sum_{i \neq j} (X_i - X_j)^2 = 2 \sum_{i < j} (X_i - X_j)^2\). \\[0.5em]
\( \text{Bias}(\hat{\theta}) = \E[\hat{\theta}] - \theta \) \dotfill Bias of an estimator \( \hat{\theta} \). \\[0.5em]
\( \text{Var}(\hat{\theta}) = \E[(\hat{\theta} - \E[\hat{\theta}])^2] \) \dotfill Variance of an estimator. \\[0.5em]
\( \text{MSE}(\hat{\theta}) = \text{Var}(\hat{\theta}) + \text{Bias}^2(\hat{\theta}) \) \dotfill Mean squared error of an estimator. \\[0.5em]
\( \vec{X} \in \mathbb{R}^n \) \dotfill Sample vector in Euclidean space. \\[0.5em]
\( \vec{1} \in \mathbb{R}^n \) \dotfill All-ones vector. \\[0.5em]
\( \vec{r} = \vec{X} - \vec{\mu} \) \dotfill Residual vector after projection onto the mean. \\[0.5em]
\( \vec{r} \in \vec{1}^\perp \) \dotfill Residual lies in the orthogonal complement of the mean vector. \\[0.5em]
\( \Gamma(k, \theta) \) \dotfill Gamma distribution with shape \( k \) and scale \( \theta \). \\[0.5em]
\( \mathcal{N}(\mu, \sigma^2) \) \dotfill Normal distribution with mean \( \mu \) and variance \( \sigma^2 \). \\[0.5em]
\( \sum X_i \) \dotfill Sum of sample values. \\[0.5em]
\( \sum X_i^2 \) \dotfill Sum of squared sample values. \\[0.5em]
\textit{Note:} The term \textit{Bariance} emphasizes variance computed from pairwise squared differences instead of deviations from the mean.

\newpage

\section{Introduction and Motivation}

Variance estimation is a foundational task in statistics and econometrics, with the sample variance being the default estimator in most applications. The unbiased version, corrected by \textbf{Bessel’s factor} (dividing by $n - 1$ rather than $n$), compensates for the loss of one degree of freedom due to pre-estimating the population mean. This correction is not just a simple algebraic trick—it admits deep geometric interpretations via orthogonal projections in $\mathbb{R}^n$ and can be derived rigorously from them.

\noindent Despite its theoretical appeal, the unbiased estimator is not always the most optimal in practice. In small samples especially, its higher variance may lead to suboptimal inference. This has led researchers to consider \textbf{shrunken estimators} that intentionally trade off a small amount of bias for a significant reduction in variance, thereby minimizing mean squared error (MSE). For example, empirical Bayes methods shrink sample variances toward a global prior, stabilizing estimation across thousands of features in genomic studies \citep{smyth2004}. Similar techniques based on James–Stein shrinkage have been explored for variance estimation in high-dimensional settings \citep{efron1975}.

\noindent Beyond the univariate case, shrinkage ideas are especially powerful in multivariate settings. In particular, shrinkage estimators for covariance matrices—such as the Ledoit–Wolf estimator \citep{ledoit2004}—have gained popularity in fields like econometrics and finance, particularly in the field of asset pricing. These estimators enhance the stability of sample covariance matrices by shrinking them toward structured targets (e.g., the identity matrix), significantly improving conditioning in high-dimensional models, which are known to perform poorly \citep{LedoitWolf2003}. This has practical relevance in the construction of variance-covariance matrices for portfolio optimization, factor models, and robust standard error estimation in large-scale regression analysis for econometric applications.

\noindent In this broader context, this paper revisits classical variance estimation and introduces a novel perspective via an alternative measure of sample dispersion based on the average squared differences between all unordered pairs in a sample. We formally define this estimator as the \textit{Bariance}, a term that reflects its construction from pairwise distances rather than deviations from a mean. It can be shown that for \textbf{mean-centered data, the \textit{Bariance} equals exactly twice the unbiased sample variance}. Moreover, a linear-time optimized formulation of the \textit{Bariance} can be derived using simple algebraic properties that avoids quadratic pairwise computation, making it both theoretically elegant and computationally efficient.

\noindent Although the pairwise difference approach has roots in classical statistics—such as U-statistics \citep{Hoeffding1948}, dissimilarity-based dispersion measures, and even the Gini coefficient \citep{Cowell2011}—the contribution here is a novel, unbiased estimator that is computationally optimized for runtime efficiency. In this respect, \textit{Bariance} bridges theoretical variance estimation with algorithmic efficiency, a consideration critical in big data contexts, real-time systems, and streaming analytics.

\noindent While computational efficiency is one of its key advantages, the \textit{Bariance} measure may also prove valuable in applied scenarios where the concept of central tendency is unstable, ill-defined, or misleading. For example, in domains such as network analysis, genomics, ordinal survey research, or clustering, statistical dispersion is often better captured through relational or pairwise structures rather than deviations from a single global mean. In such contexts, the \textit{Bariance} shares conceptual kinship with the Gini coefficient, which also operates on pairwise differences but in a distributional inequality framework. Unlike Gini, however, \textit{Bariance} preserves unbiasedness for variance estimation under i.i.d.\ sampling and scales naturally in high-dimensional or streaming environments. These features make it particularly attractive for modern applications in unsupervised learning, robust statistics, and high-throughput data pipelines—where traditional variance measures may either fail or become computationally prohibitive.

\noindent Through an empirical simulation study, I demonstrate that this \textbf{optimized unbiased sample variance estimator} remains unbiased and improves runtime. The simulated empirical runtimes section includes confidence intervals, hardware specifications, seed initialization, and multiple replications, thereby addressing robustness, reproducibility, and statistical reliability. Furthermore, Appendix C replicates the result in a local hardware environment using an alternative high-level programming language. We then revisit the controversial idea—advocated by Rosenthal \citep{rosenthal2015}—that dividing by $n$ (rather than $n - 1$) may yield lower-MSE variance estimators in practice, especially when unbiasedness is not strictly required.

\noindent To sum up, the \textit{Bariance} framework bridges computational efficiency with applied relevance, offering a theoretically grounded yet practically flexible alternative to traditional variance estimators. This paper thus aims to bridge classical econometric and statistical theory with modern considerations of efficiency, robustness, and computational scalability, while highlighting the often underestimated choices in estimator design or usage.

\newpage

\section{Definitions and Setup}

Let \( X_1, X_2, \dots, X_n \in \mathbb{R} \) be i.i.d.\ random variables with:
\[
\E[X_i] = \mu, \qquad \Var(X_i) = \sigma^2
\]

\noindent Define the sample mean and biased/unbiased variance estimates:
\[
\bar{X} := \frac{1}{n} \sum_{i=1}^n X_i, \quad
S^2 := \frac{1}{n} \sum_{i=1}^n (X_i - \bar{X})^2, \quad
\hat{S}^2 := \frac{1}{n-1} \sum_{i=1}^n (X_i - \bar{X})^2
\]

\section{Derivation of Bias and Bessel’s Correction}

An estimator \( \hat{\theta} \) for a parameter \( \theta \) is called \textbf{unbiased} if its expected value equals the true value:
\[
\E[\hat{\theta}] = \theta
\]

\noindent The normal $n$-based \textbf{sample variance with denominator \( n \)} is defined as:

\[
S^2 := \frac{1}{n} \sum_{i=1}^n (X_i - \bar{X})^2
\]

\noindent We aim to compute \( \E[S^2] \), the expected value of this estimator, to show that it is biased.\newline 

\noindent \textbf{Expand the squared deviations:}

\[
\sum_{i=1}^n (X_i - \bar{X})^2 \equiv \sum_{i=1}^n X_i^2 - n\bar{X}^2
\]

\noindent Thus:

\[
S^2 = \frac{1}{n} \left( \sum_{i=1}^n X_i^2 - n\bar{X}^2 \right)
= \frac{1}{n} \sum X_i^2 - \bar{X}^2
\]

\noindent \textbf{Then, take expectation of \( S^2 \).} By linearity of expectation to each term:

\[
\E[S^2] = \frac{1}{n} \sum_{i=1}^n \E[X_i^2] - \E[\bar{X}^2]
\]

\noindent \textbf{Compute \( \E[X_i^2] \)}. Using the known identity:

\[
\E[X_i^2] \equiv \Var(X_i) + (\E[X_i])^2 = \sigma^2 + \mu^2
\]

\noindent So, the $n$ cancels out, eventually:

\[
\frac{1}{n} \sum_{i=1}^n \E[X_i^2] = \frac{1}{n} \cdot n(\mu^2 + \sigma^2) = \mu^2 + \sigma^2
\]

\noindent \textbf{Compute \( \E[\bar{X}^2] \)}. Recall that:

\[
\bar{X} = \frac{1}{n} \sum_{i=1}^n X_i
\quad \Rightarrow \quad
\E[\bar{X}] = \mu, \quad
\Var(\bar{X}) = \frac{\sigma^2}{n}
\]

\noindent Thus:

\[
\E[\bar{X}^2] = \Var(\bar{X}) + (\E[\bar{X}])^2 
= \frac{\sigma^2}{n} + \mu^2
\]

\noindent \textbf{Combining both terms now}:

\[
\E[S^2] = (\mu^2 + \sigma^2) - \left( \mu^2 + \frac{\sigma^2}{n} \right)
= \sigma^2 - \frac{\sigma^2}{n}
= \left( \frac{n - 1}{n} \right) \sigma^2
\]

\[
\boxed{\E[S^2] = \frac{n - 1}{n} \sigma^2}
\]

\noindent This shows that the estimator \( S^2 \) is \textbf{biased}, underestimating the population variance \( \sigma^2 \), because the denominator is larger than the numerator.

\noindent \textbf{Bessel’s Correction.} To correct the bias, we define the \textbf{unbiased sample variance} as:

\[
\hat{S}^2 := \frac{1}{n - 1} \sum_{i=1}^n (X_i - \bar{X})^2
\Rightarrow
\E[\hat{S}^2] = \sigma^2
\]

\[
\boxed{
\E[\hat{S}^2] = \sigma^2 \quad \text{(unbiased)}
}
\]

\noindent This is known as \textbf{Bessel’s correction} — using \( n-1 \) instead of \( n \) in the denominator compensates for the loss of one degree of freedom from estimating the mean \( \mu \) with \( \bar{X} \).

\newpage

\section{Geometric Interpretation of Estimated Variance and $n-1$ Degrees of Freedom}

\subsection*{Orthogonal Decomposition}

Let $\vec{X} := \begin{bmatrix} X_1 & \dots & X_n \end{bmatrix}^\top \in \mathbb{R}^n$. Define the mean
\[
\bar{X} = \frac{1}{n} \sum_{i=1}^n X_i, \quad \vec{\mu} = \bar{X} \cdot \vec{1}.
\]
Then
\[
\vec{X} = \vec{\mu} + \vec{r}, \quad \vec{r} := \vec{X} - \vec{\mu}, \quad \vec{r} \in \vec{1}^\perp.
\]
Indeed, $\vec{1}^\top \vec{r} = \sum_{i=1}^n (X_i - \bar{X}) = 0$.

\begin{center}
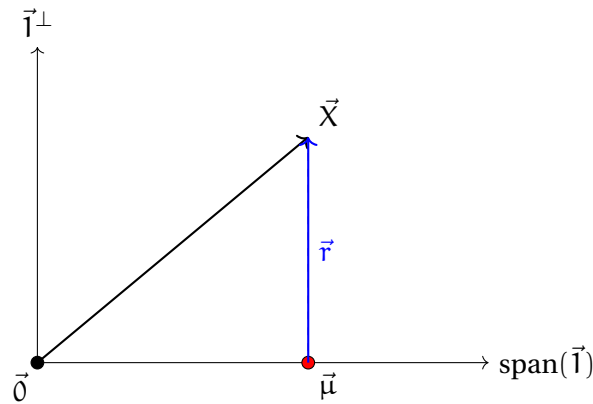

\begin{tikzpicture}[scale=1.2]
\draw[->] (0,0) -- (5,0) node[right] {$\operatorname{span}(\vec{1})$};
\draw[->] (0,0) -- (0,3.5) node[above] {$\vec{1}^\perp$};

\coordinate (X) at (3,2.5);
\coordinate (mu) at (3,0);

\draw[->, thick] (0,0) -- (X) node[above right] {$\vec{X}$};
\draw[dashed] (X) -- (mu);
\draw[fill=red] (mu) circle (2pt) node[below right] {$\vec{\mu}$};
\draw[->, thick, blue] (mu) -- (X) node[midway, right] {$\vec{r}$};
\draw[fill=black] (0,0) circle (2pt) node[below left] {$\vec{0}$};
\end{tikzpicture}

\captionof{figure}{Orthogonal decomposition of $\vec{X}$}
\end{center}

\subsection*{Dimension and Degrees of Freedom}

\begin{align*}
\vec{X} &\in \mathbb{R}^n \\
\vec{\mu} &\in \operatorname{span}(\vec{1}),\quad \dim = 1 \\
\vec{r} &\in \vec{1}^\perp,\quad \dim = n - 1 \\
\Rightarrow\quad \text{DoF} &= n - 1
\end{align*}

\subsection*{Unbiased Sample Variance}

\[
s^2 = \frac{1}{n - 1} \| \vec{r} \|^2 = \frac{1}{n - 1} \sum_{i=1}^n (X_i - \bar{X})^2
\]

\subsection*{Common Application using Orthogonal Decomposition for Dimensionality reduction: Principal Component Analysis (PCA)}

Let $\mathbf{X} \in \mathbb{R}^{n \times p}$ be a data matrix with rows as observations and columns as variables. Let $\tilde{X} := X - \vec{1} \bar{x}^\top$, where $\bar{x} = \frac{1}{n} X^\top \vec{1} \in \mathbb{R}^p$.

PCA seeks orthonormal vectors $\vec{v}_i \in \mathbb{R}^p$ satisfying:
\[
\tilde{X}^\top \tilde{X} \vec{v}_i = \lambda_i \vec{v}_i, \quad i = 1, \dots, p.
\]
Then $\vec{z}_i := \tilde{X} \vec{v}_i$ are commonly known as $i$th principal components ($PC_i$), with variances $\operatorname{Var}(\vec{z}_i) = \lambda_i / (n - 1)$.

\section{Introducing the \textit{Bariance} and an optimized linear Estimator}

We define the \textbf{\textit{Bariance}} of a sample \( \{X_1, X_2, \dots, X_n\} \) as the average squared difference over all unordered pairs:

\[
\text{\textit{Bariance}} := \frac{1}{n(n-1)} \sum_{i \ne j} (X_i - X_j)^2
\]

\noindent The term \textit{Bariance} is simply chosen to emphasize the estimator's foundation on pairwise \textbf{b}etween-sample v\textbf{ariance} rather than deviations from the mean, highlighting its construction from pairwise squared differences. This can also be interpreted as the average squared length of all edges in a complete graph on the sample points.

\noindent \textbf{We begin by expanding the inner squared difference}:

\[
(X_i - X_j)^2 = X_i^2 - 2X_i X_j + X_j^2
\]

\noindent Summing over all distinct \( i \ne j \):

\[
\sum_{i \ne j} (X_i - X_j)^2 
= \sum_{i \ne j} (X_i^2 + X_j^2 - 2X_i X_j)
\]

\noindent We split this into three terms:

\[
= \sum_{i \ne j} X_i^2 + \sum_{i \ne j} X_j^2 - 2 \sum_{i \ne j} X_i X_j
\]

\noindent Note the following observations:
- For fixed \( i \), there are \( n-1 \) values of \( j \ne i \), so:
  \[
  \sum_{i \ne j} X_i^2 = (n - 1) \sum_{i=1}^n X_i^2
  \]
  Similarly, \( \sum_{i \ne j} X_j^2 = (n - 1) \sum_{j=1}^n X_j^2 \)

\noindent So the first two terms become:
\[
\sum_{i \ne j} X_i^2 + \sum_{i \ne j} X_j^2 = 2(n-1) \sum_{i=1}^n X_i^2
\]

\noindent Now consider the double sum:
\[
\sum_{i \ne j} X_i X_j = \left( \sum_{i=1}^n \sum_{j=1}^n X_i X_j \right) - \sum_{i=1}^n X_i^2
= \left( \sum X_i \right)^2 - \sum X_i^2
\]

\noindent Combine:

\[
\sum_{i \ne j} (X_i - X_j)^2
= 2(n-1) \sum X_i^2 - 2\left( \left( \sum X_i \right)^2 - \sum X_i^2 \right)
\]

\[
= 2(n-1) \sum X_i^2 - 2 \left( \sum X_i \right)^2 + 2 \sum X_i^2
= 2n \sum X_i^2 - 2 \left( \sum X_i \right)^2
\]

\noindent \textbf{Substitute back into the \textit{Bariance} formula}. Now divide by \( n(n-1) \):

\[
\text{\textit{Bariance}} = \frac{1}{n(n-1)} \sum_{i \ne j} (X_i - X_j)^2
\equiv \frac{2n}{n(n-1)} \sum X_i^2 - \frac{2}{n(n-1)} \left( \sum X_i \right)^2.
\]

\noindent For empirical verification of this algebraic identity see Appendix A.

\[
\boxed{
\text{\textit{Bariance}}_{\text{opt}} := \frac{2n}{n(n - 1)} \sum X_i^2 - \frac{2}{n(n - 1)} \left( \sum X_i \right)^2.
}
\]

\subsubsection{In the Case Of Mean-centered data}

\noindent If the data is centered, i.e., \( \sum X_i = 0 \), then:

\[
\text{\textit{Bariance}} = \frac{2n}{n(n - 1)} \sum X_i^2 = \frac{2}{n - 1} \sum X_i^2.
\]

\noindent We now relate this to the unbiased sample variance estimator:

\[
\hat{S}^2 = \frac{1}{n - 1} \sum_{i=1}^n (X_i - \bar{X})^2
= \frac{1}{n - 1} \sum X_i^2 \quad \text{(since } \bar{X} = 0 \text{)}.
\]

\noindent Therefore, the following equality holds for the defined ``\textit{Bariance}``:

\[
\boxed{
\text{\textit{Bariance}} = 2 \cdot \hat{S}^2.
}
\]

\noindent This result shows that \textit{Bariance} represents twice the unbiased sample variance when the sample is mean-centered. It provides an elegant \textbf{pairwise perspective} on variance: instead of summing squared deviations from a central value, we sum squared differences between all pairs and average, regardless of the reference point within the sample.

\subsection{Properties of the \textit{Bariance} and Estimator Comparison}

Let \( \theta := \sigma^2 = 1 \). Then:

\[
\begin{aligned}
\E[\hat{S}^2] &= \theta, \\
\E[\text{\textit{Bariance}}] &= 2\theta, \\
\text{Bias}(\hat{S}^2) &= 0, \\
\text{Bias}(\text{\textit{Bariance}}) &= \theta, \\
\Var(\hat{S}^2) &= \frac{2\theta^2}{n - 1}, \\
\Var(\text{\textit{Bariance}}) &= 4 \cdot \Var(\hat{S}^2), \\
\text{MSE}(\hat{S}^2) &= \Var(\hat{S}^2), \\
\text{MSE}(\text{\textit{Bariance}}) &= 4 \cdot \Var(\hat{S}^2) + \theta^2.
\end{aligned}
\]

\noindent \subsubsection*{Subsequent numerical verification of theoretical relationships with \( n = 100 \) and \( \tau = 1000 \)}
\noindent Numerically (for \( n = 100 \), \( \mathcal{N}(0,1) \), \( \tau=1000 \)):

\vspace{0.5em}

\begin{center}
\begin{tabular}{|l|c|c|c|c|}
\hline
\textbf{Estimator} & \textbf{Point Estimate} & \textbf{Bias} & \textbf{Variance} & \textbf{MSE} \\
\hline
Unbiased sample variance estimator & 1.00091 & 0.00091 & 0.02156 & 0.02151 \\
\textit{Bariance} & 2.00181 & 1.00181 & 0.08625 & 1.08968 \\
\hline
\end{tabular}
\end{center}

\[
\Var(\hat{S}^2) = 0.01988, \quad
\Var(2 \cdot \hat{S}^2) = 0.07951,
\]
\[
\text{MSE}(\hat{S}^2) = 0.02151, \quad
\text{MSE}(2 \cdot \hat{S}^2) = 1.0832,
\]
\[
4 \cdot \Var(\hat{S}^2) = 0.07951 \quad \Rightarrow \quad
\text{Var identity approximately holds},
\]
\[
4 \cdot \Var(\hat{S}^2) + \theta^2 = 0.07951 + 1 = 1.07951 \quad \Rightarrow \quad
\text{MSE identity approximately holds}.
\]
\[
\boxed{
\begin{aligned}
\text{Bias}(\text{\textit{Bariance}}) &:= \theta\\
Var(Bariance) &:= 4 \cdot \Var(\hat{S}^2), \\
\text{MSE}(Bariance) &:= 4 \cdot \Var(\hat{S}^2) + \theta^2.
\end{aligned}
}
\]

\noindent \textbf{Summary}:

\begin{itemize}
    \item \( \hat{S}^2 \): unbiased, lower MSE, standard estimator.
    \item \textit{Bariance}: biased (by \( +\theta \)), higher MSE due to both variance inflation and squared bias.
    \item Useful relation: \( \text{\textit{Bariance}} = 2 \cdot \hat{S}^2 \). (Because: $\theta$ = ($\theta$ + $\theta$)/2)
\end{itemize}

\subsection{A Graph-Theoretic View of \textit{Bariance}}

\begin{center}
\begin{tikzpicture}[scale=1.2]
\coordinate (X1) at (1,1);
\coordinate (X2) at (2.5,2.2);
\coordinate (X3) at (4,1);
\coordinate (X4) at (3.5,0);
\coordinate (X5) at (2,0.3);

\foreach \pt/\lbl in {X1/$X_1$, X2/$X_2$, X3/$X_3$, X4/$X_4$, X5/$X_5$} {
  \draw[fill=black] (\pt) circle (1.5pt);
  \node at ($( \pt ) + (0.2, 0.2)$) {\lbl};
}

\foreach \i/\j in {X1/X2, X1/X3, X1/X4, X1/X5,
                   X2/X3, X2/X4, X2/X5,
                   X3/X4, X3/X5,
                   X4/X5}
  \draw[gray!50, dashed] (\i) -- (\j);
\end{tikzpicture}
\captionof{figure}{Complete graph over sample points. Each edge represents a pairwise squared difference that contributes to the \textit{Bariance}.}
\end{center}

Each dashed edge on the graph corresponds to a pair \( (X_i, X_j) \), and the squared difference \( (X_i - X_j)^2 \) can be viewed as an edge weight. Since the graph is fully connected (a complete graph \( G_n \)), there are \( \binom{n}{2} \) such edges.

\noindent From a graph-theoretic standpoint, \textit{Bariance} is the \textbf{average squared edge length} of the complete weighted graph over the sample. In this context:

\noindent - \textbf{Nodes} = observations \( X_i \) \\
- \textbf{Edges} = pairwise differences \\
- \textbf{Weights} = squared differences \( (X_i - X_j)^2 \)

\noindent This perspective connects naturally to \textbf{distance-based dispersion measures} in statistical graph theory, including energy statistics and certain U-statistics. It also provides an intuitive, coordinate-free alternative to the variance's reliance on a central location (i.e., the mean).

\subsection{Deviation from Mean (Variance) vs. Pairwise Differences (\textit{Bariance} Components)}

\begin{center}
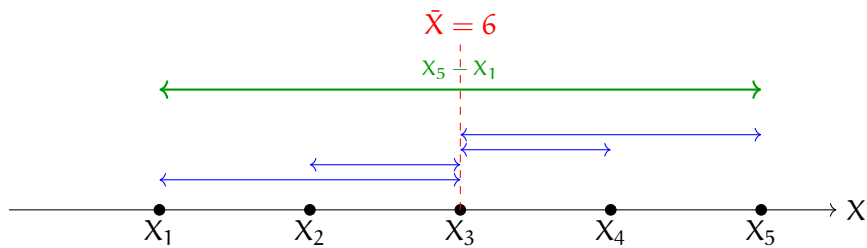

\begin{tikzpicture}
\draw[->] (0,0) -- (11,0) node[right] {$X$};

\foreach \x/\label in {2/$X_1$, 4/$X_2$, 6/$X_3$, 8/$X_4$, 10/$X_5$}
  \draw[fill=black] (\x,0) circle (2pt) node[below] {\label};

\draw[dashed, red] (6,0) -- (6,2.2) node[above] {\textcolor{red}{$\bar{X} = 6$}};
\foreach \x/\h in {2/0.4,4/0.6,8/0.8,10/1.0}
  \draw[<->, blue] (\x,\h) -- (6,\h);

\draw[<->, green!60!black, thick] (2,1.6) -- (10,1.6) node[midway, above] {\scriptsize $X_5 - X_1$};
\end{tikzpicture}
\captionof{figure}{Blue: variance (mean-deviation, \(n-1\) degrees of freedom adjustment). Green: pairwise distance = a \textit{Bariance} component.}
\end{center}

\subsection{The Pairwise Difference Grid}

\begin{center}
\begin{tikzpicture}
\foreach \i in {0,...,4} {
    \foreach \j in {0,...,4} {
        \pgfmathsetmacro\val{(2*\i - 2*\j)^2}
        \node at (\j, -\i) {\val};
    }
}
\draw[step=1cm, transparent, very thin] (0,-4) grid (4,0);
\foreach \i/\val in {0/2,1/4,2/6,3/8,4/10} {
    \node at (-1.5,-\i) {\val};
    \node at (\i, 0.5) {\val};
}
\end{tikzpicture}
\captionof{figure}{Symmetric matrix of squared pairwise differences \( (X_i - X_j)^2 \) for \( X = \{2, 4, 6, 8, 10\} \). Diagonal elements are zero (self-distances).}
\end{center}

\noindent This matrix visualizes all pairwise squared differences between elements of \( X \). Its key properties:

\noindent - \textbf{Symmetr}y: \( (X_i - X_j)^2 = (X_j - X_i)^2 \), so the grid is symmetric across the main diagonal. \\
- \textbf{Diagonal}: Always zero, since \( (X_i - X_i)^2 = 0 \). \\
- \textbf{Off-diagonal structure}: Each non-zero entry contributes to the total in the \textit{Bariance} calculation.

\noindent This grid-based view reinforces how the redundancy in the pairwise differences allows us to simplify the computation algebraically. Instead of evaluating all \( n(n - 1) \) pairs, we can summarize the matrix using aggregate row/column sums, which enables the linear-time algebraically optimized derived formula for the \textit{Bariance}:

\[
\text{\textit{Bariance}}_{\text{opt}} = \frac{2n}{n(n-1)} \sum X_i^2 - \frac{2}{n(n-1)} \left( \sum X_i \right)^2.
\]

\noindent Thus, symmetry is not just a visual feature—it underpins the algebraic transformation that reduces the quadratic computational complexity to linear.

\subsection{Numerical Verification of \textit{Bariance}-Estimator properties using Gamma-distributed Data}

Let \( X \sim \Gamma(k = 2, \theta = 2) \Rightarrow \E[X] = 4, \quad \Var(X) = 8 \), Using \( n = 100 \), \( \tau = 1000 \) we obtain:

\subsubsection*{Summary Table of Empirical \textit{Bariance} Point Estimator Performance}

\begin{center}
\begin{tabular}{|l|c|c|c|c|}
\hline
\textbf{Estimator} & \textbf{Mean} & \textbf{Bias} & \textbf{Variance} & \textbf{MSE} \\
\hline
Unbiased Sample Variance Estimator & 8.00087 & 0.00087 & 2.92574 & 2.92281 \\
\( \text{\textit{Bariance}} \) & 16.00174 & 8.00174 & 11.70295 & 75.71907 \\
\hline
\end{tabular}
\end{center}

\subsubsection*{Numerical Verification of Theoretical Relationships}

Let \( \sigma^2 := 8 \Rightarrow \sigma^4 = 64 \):

\[
\Var(\hat{S}^2) = 2.92574, \quad
\Var(\text{\textit{Bariance}}) = 11.70295,
\]

\[
\text{Empirical MSE}(\text{\textit{Bariance}}) = 75.71907,
\]

We numerically verify:

\[
\Var(\text{\textit{Bariance}}) \approx 4 \cdot \Var(\hat{S}^2)
\quad \Rightarrow \quad
11.70295 \approx 4 \cdot 2.92574 = 11.70296,
\]

\[
\text{MSE}(\text{\textit{Bariance}}) \approx 4 \cdot \Var(\hat{S}^2) + \sigma^4
\quad \Rightarrow \quad
75.71907 \approx 4 \cdot 2.92574 + 64 = 75.70295,
\]

\[
\text{Equalities hold numerically with high accuracy:} \quad
\begin{cases}
\Var(\text{\textit{Bariance}}) = 4 \cdot \Var(\hat{S}^2), \\
\text{MSE}(\text{\textit{Bariance}}) = 4 \cdot \Var(\hat{S}^2) + \sigma^4.
\end{cases}
\]

\clearpage
\newpage

\section{Discussion: Should We Just Divide by \( n \)?}

\noindent Rosenthal \citep{rosenthal2015} argues that using \( n \) instead of \( n - 1 \) may lead to a \textbf{smaller mean squared error (MSE)} — especially when teaching or in practical settings.

\noindent He shows that while dividing by \( n - 1 \) yields an unbiased estimator, this may come at the cost of increased variance. In some cases, a biased but lower-MSE estimator using \( n \) is preferable:
\begin{quote}
“...a smaller, shrunken, biased estimator actually reduces the MSE...” — \citep{rosenthal2015}
\end{quote}

\noindent This introduces another viewpoint: unbiasedness isn’t always the ultimate goal — \textbf{minimizing error in practice often is}.  

\noindent From a theoretical perspective, unbiasedness ensures that the expected value of the estimator exactly matches the true population variance. However, unbiasedness alone does not guarantee minimal estimation error in finite samples. In fact, particularly when sample sizes are small, the variance of the unbiased estimator can be relatively large, which may lead to unstable or inefficient estimates. Allowing a small bias can reduce this variance enough to produce a lower overall mean squared error, which combines both bias and variance in a single measure of estimator quality \citep{Casella2002, Shao2003}. 

\noindent Practically, this trade-off becomes important in many applied contexts, such as when the variance estimate is an intermediate quantity used for further modeling or prediction. Here, reducing total estimation error takes precedence over strict unbiasedness. Additionally, computational simplicity and pedagogical clarity sometimes favor the \( n \)-denominator estimator, which is easier to understand and implement, especially in introductory settings.

\noindent To illustrate, consider the case of \( n = 5 \) observations drawn from a population with true variance \( \sigma^2 = 10 \). The biased estimator, which divides by \( n = 5 \), underestimates the variance as \( 8 \), while the unbiased estimator, dividing by \( n - 1 = 4 \), correctly yields \( 10 \). However, when mean squared error is calculated, which accounts for both bias and variance, the biased estimator can have smaller total error due to its reduced variance. Rosenthal explicitly notes that the variance of the biased estimator is often less than that of the unbiased one, which explains this result.

\noindent This nuanced understanding clarifies why and when the classical insistence on unbiasedness may be relaxed in favor of better finite-sample performance and practical utility. Furthermore, more generalized estimators with a denominator parameter \( a > 0 \), defined as
\[
\hat{\sigma}^2_a := \frac{1}{a} \sum_{i=1}^n (X_i - \bar{X})^2,
\]
can be analyzed for optimality in terms of mean squared error. As derived in Appendix~B, the MSE-minimizing denominator is approximately \( a^* \approx n + 1 \), suggesting that neither the classical unbiased divisor \( n - 1 \) nor the biased \( n \) divisor are necessarily optimal from an MSE standpoint.

\noindent In sum, relaxing unbiasedness for variance estimation is a principled and context-dependent choice motivated by the bias-variance trade-off, sample size considerations, and the ultimate goals of estimation. This perspective complements classical theory and better reflects the realities of applied statistical practice.

\section{A Simulation Study: Bias$^2$, Variance, and MSE Across Denominator Values}

We consider the family of estimators for the population variance \( \sigma^2 \):

\[
\hat{\sigma}^2_a := \frac{1}{a} \sum_{i=1}^n (X_i - \bar{X})^2
\quad \text{for varying } a > 0
\]

The simulation is carried out with the following parameters:
\begin{itemize}
  \item Sample size: \( n = 5 \)
  \item True variance: \( \sigma^2 = 10 \)
  \item Distribution: \( X_i \sim \mathcal{N}(0, \sigma^2) \)
  \item Number of simulations: \( 100{,}000 \)
\end{itemize}

\noindent For each value of \( a \in [3.5, 8.5] \) (in increments of 0.5), we compute the following empirically:
\begin{align*}
\text{Bias}(\hat{\sigma}^2_a) &= \mathbb{E}[\hat{\sigma}^2_a] - \sigma^2 \\
\text{Bias}^2 &= \left(\mathbb{E}[\hat{\sigma}^2_a] - \sigma^2\right)^2 \\
\text{Variance} &= \text{Var}[\hat{\sigma}^2_a] \\
\text{MSE} &= \text{Bias}^2 + \text{Variance}
\end{align*}

\subsection*{Empirical Results}

\begin{table}[h]
\centering
\caption{Empirical Bias$^2$, Variance, and MSE with 95\% bootstrapped confidence intervals (200 resamples, seed=42). Bold rows indicate $a = n-1$, $n$, and $n+1$. Hardware: 1GB RAM, Python 3.11. For a theoretical derivation of the MSE-minimized variance estimator, see Appendix B.}
\label{tab:empirical-mse-ci-bootstrap}
\begin{tabular}{rccc}
\toprule
$a$ & Bias$^2$ [CI] & Variance [CI] & MSE [CI] \\
\midrule
{3.5} & {1.98 [1.97, 2.03]} & {61.99 [61.85, 62.19]} & {63.96 [63.83, 64.21]} \\
\textbf{4.0} & \textbf{0.00 [0.00, 0.01]} & \textbf{47.46 [47.26, 47.51]} & \textbf{47.46 [47.26, 47.52]} \\
{4.5} & {1.27 [1.26, 1.30]} & {37.50 [37.34, 37.57]} & {38.77 [38.63, 38.84]} \\
\textbf{5.0} & \textbf{4.06 [4.02, 4.08]} & \textbf{30.37 [30.29, 30.47]} & \textbf{34.44 [34.36, 34.50]} \\
{5.5} & {7.52 [7.48, 7.54]} & {25.10 [25.07, 25.20]} & {32.62 [32.59, 32.70]} \\
\textbf{6.0} & \textbf{11.20 [11.16, 11.24]} & \textbf{21.09 [21.05, 21.17]} & \textbf{32.29 [32.27, 32.36]} \\
{6.5} & {14.89 [14.81, 14.90]} & {17.97 [17.92, 18.03]} & {32.86 [32.79, 32.87]} \\
{7.0} & {18.46 [18.46, 18.56]} & {15.50 [15.42, 15.51]} & {33.96 [33.94, 34.01]} \\
{7.5} & {21.88 [21.85, 21.94]} & {13.50 [13.48, 13.55]} & {35.37 [35.37, 35.44]} \\
{8.0} & {25.10 [25.06, 25.17]} & {11.86 [11.82, 11.89]} & {36.96 [36.93, 37.01]} \\
{8.5} & {28.13 [28.09, 28.20]} & {10.51 [10.45, 10.52]} & {38.64 [38.59, 38.67]} \\
\bottomrule
\end{tabular}
\end{table}

\begin{figure}[h]
\centering
\begin{tikzpicture}
\begin{axis}[
    width=14cm,
    height=12cm,
    xlabel={Denominator $a$},
    ylabel={Value},
    title={MSE, Bias$^2$, and Variance of Variance Estimator},
    legend style={at={(0.01,0.98)}, anchor=north west},
    grid=major,
    ymin=0, ymax=50
]

\addplot+[mark=*, thick, blue] coordinates {
(3.5, 63.96) (4.0, 47.46) (4.5, 38.77) (5.0, 34.44)
(5.5, 32.62) (6.0, 32.29) (6.5, 32.86) (7.0, 33.96)
(7.5, 35.37) (8.0, 36.96) (8.5, 38.64)
};
\addlegendentry{MSE}

\addplot+[mark=triangle*, thick, red, dashed] coordinates {
(3.5, 1.98) (4.0, 0.00) (4.5, 1.27) (5.0, 4.06)
(5.5, 7.52) (6.0, 11.20) (6.5, 14.89) (7.0, 18.46)
(7.5, 21.88) (8.0, 25.10) (8.5, 28.13)
};
\addlegendentry{Bias$^2$}

\addplot+[mark=square*, thick, green!60!black, dashed] coordinates {
(3.5, 61.99) (4.0, 47.46) (4.5, 37.50) (5.0, 30.37)
(5.5, 25.10) (6.0, 21.09) (6.5, 17.97) (7.0, 15.50)
(7.5, 13.50) (8.0, 11.86) (8.5, 10.51)
};
\addlegendentry{Variance}

\addplot[dotted, thick, black] coordinates {(4, 0) (4, 70)};
\addlegendentry{Divide by $n{-}1$}

\addplot[dotted, thick, gray] coordinates {(5, 0) (5, 70)};
\addlegendentry{Divide by $n$}

\addplot[dotted, thick, gray!50!white] coordinates {(6, 0) (6, 70)};
\addlegendentry{Divide by $n{+}1$}

\end{axis}
\end{tikzpicture}
\caption{Empirical MSE, Bias$^2$, and Variance of the sample variance estimator for $a \in [3.5, 8.5]$ and $n=5$ using 10,000 simulations and 200 bootstraps. Minimum MSE occurs between $a=5.5$ and $a=6.5$. For a theoretical derivation of the MSE-minimized variance estimator, see Appendix B.}
\end{figure}
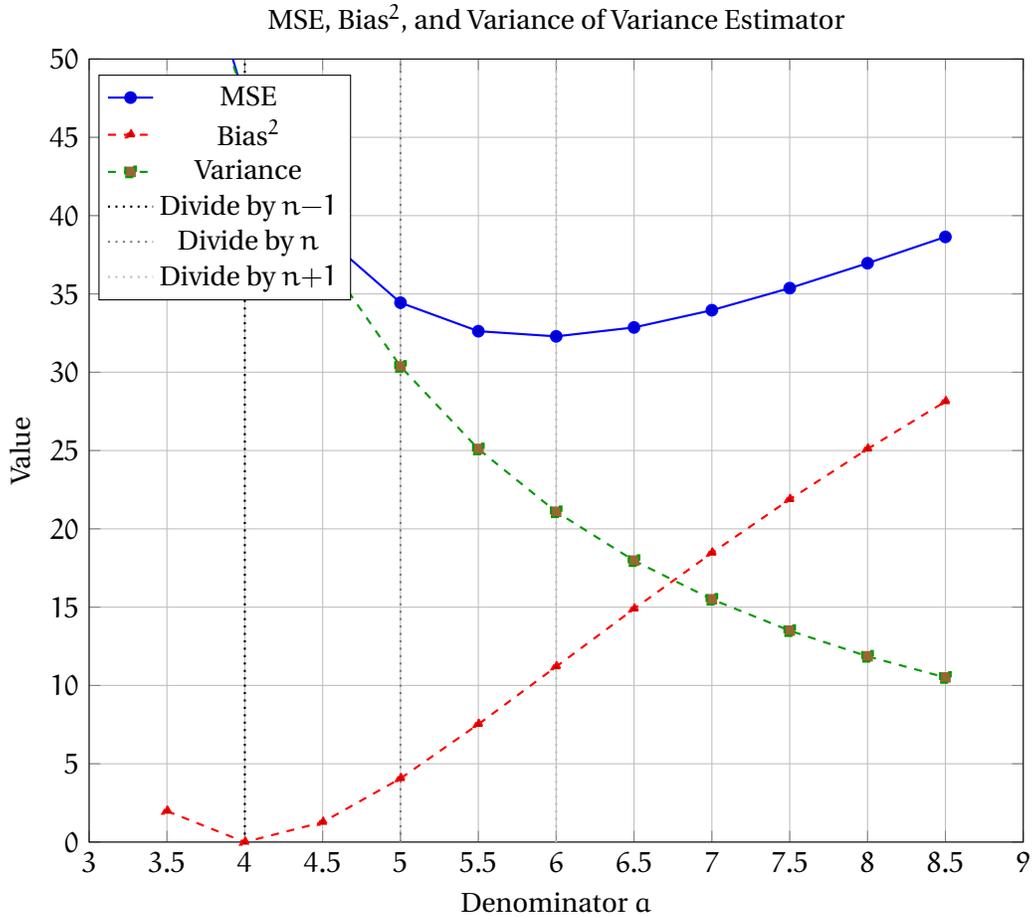

\clearpage
\newpage

\section{On Computational Complexity of Variance \textit{Bariance} Estimators and Optimization}

Let \( X := \{X_1, X_2, \dots, X_n\} \subset \mathbb{R} \) be a sample of size \( n \).

\begin{table}[h!]
\centering
\caption{Computational complexity of variance and \textit{Bariance} estimators with explanation}
\label{tab:complexity-detail}
\begin{tabular}{p{5.5cm} p{7.5cm} p{2.4cm}}
\toprule
\textbf{Estimator} & \textbf{Operations} & \textbf{Complexity} \\
\midrule

\textbf{Biased Variance} \\
\( S^2 = \frac{1}{n} \sum (X_i - \bar{X})^2 \) 
& 
\begin{itemize}[leftmargin=*]
\item 1 pass to compute mean \( \bar{X} \)
\item 1 pass to compute squared deviations
\item Total: 2 linear scans
\item For \( n = 5 \): 5 additions, 5 subtractions, 5 squarings
\end{itemize}
& 
\( \mathcal{O}(n) \) \\

\addlinespace[1ex]

\textbf{Unbiased Variance} \\
\( \hat{S}^2 = \frac{1}{n - 1} \sum (X_i - \bar{X})^2 \)
& 
Same steps as biased estimator; only the divisor differs. 
\newline
No added computation.
&
\( \mathcal{O}(n) \) \\

\addlinespace[1ex]

\textbf{\textit{Bariance} (Naïve)} \\
\( \frac{1}{n(n - 1)} \sum_{i \ne j} (X_i - X_j)^2 \)
& 
\begin{itemize}[leftmargin=*]
\item All \( n(n-1) \) ordered pairs evaluated
\item Each requires subtraction + squaring
\item For \( n = 5 \): \( 5 \times 4 = 20 \) pairs
\item Cost grows quadratically with sample size
\end{itemize}
&
\( \mathcal{O}(n^2) \) \\

\addlinespace[1ex]

\textbf{\textit{Bariance} (Optimized)} \\
\( \frac{2n}{n(n-1)} \sum X_i^2 - \frac{2}{n(n-1)} \left( \sum X_i \right)^2 \)
& 
\begin{itemize}[leftmargin=*]
\item Uses 2 scalar sums: \( \sum X_i \), \( \sum X_i^2 \)
\item Each computed in 1 pass
\item For \( n = 5 \): 5 additions, 5 squarings
\end{itemize}
& 
\( \mathcal{O}(n) \) \\
\bottomrule
\end{tabular}
\end{table}

\newpage
\subsection{Computational Complexity Comparison with Numerical Illustration}

We compare the computational cost of the biased variance, unbiased variance, and \textit{Bariance} estimators using both theoretical analysis and a numerical example for \( n = 5 \).

\subsection*{Example: \( X = \{1, 3, 5, 7, 9\} \)}

\textbf{Mean:}
\[
\bar{X} = \frac{1 + 3 + 5 + 7 + 9}{5} = \frac{25}{5} = 5
\]

\noindent \textbf{Biased Variance:}
\[
S^2 = \frac{1}{5} \sum (X_i - \bar{X})^2 
= \frac{1}{5} [(1-5)^2 + (3-5)^2 + (5-5)^2 + (7-5)^2 + (9-5)^2] 
= \frac{1}{5} [16 + 4 + 0 + 4 + 16] 
= \frac{40}{5} = 8
\]

\noindent \textbf{Unbiased Variance:}
\[
\hat{S}^2 = \frac{1}{4} \sum (X_i - \bar{X})^2 = \frac{40}{4} = 10
\]

\noindent \textbf{Naïve \textit{Bariance}:}
\[
\sum_{i < j} (X_i - X_j)^2 
= (3-1)^2 + (5-1)^2 + (7-1)^2 + (9-1)^2 + (5-3)^2 + (7-3)^2 + (9-3)^2 + (7-5)^2 + (9-5)^2 + (9-7)^2
\]
\[
= 4 + 16 + 36 + 64 + 4 + 16 + 36 + 4 + 16 + 4 = 200
\]

\[
\text{\textit{Bariance}} = \frac{2 \cdot 200}{5 \cdot 4} = \frac{400}{20} = 20
\]

\noindent \textbf{Optimized \textit{Bariance}:}

\[
\sum X_i = 1 + 3 + 5 + 7 + 9 = 25, \quad \sum X_i^2 = 1^2 + 3^2 + 5^2 + 7^2 + 9^2 = 165
\]

\[
\text{\textit{Bariance}} = \frac{2n}{n(n-1)} \sum X_i^2 - \frac{2}{n(n-1)} \left(\sum X_i\right)^2 = \frac{2 \cdot 5}{20} \cdot 165 - \frac{2}{20} \cdot 625 = \frac{1650}{20} - \frac{1250}{20} = 82.5 - 62.5 = 20
\]

\noindent Thus, all estimators yield consistent results, confirming their correctness. However, their computational complexity differs:

\begin{itemize}
  \item Biased/Unbiased Variance: \(\mathcal{O}(n)\)
  \item Naïve \textit{Bariance}: \(\mathcal{O}(n^2)\)
  \item Optimized \textit{Bariance}: \(\mathcal{O}(n)\)
\end{itemize}

\noindent The optimized \textit{Bariance} offers the same result as the naïve form but with significantly reduced computational cost, making it efficient for large-scale applications.

\clearpage
\newpage

\section{On Empirical Runtime of \textit{Bariance} Estimators}

\noindent To evaluate the practical performance of variance and \textit{Bariance} estimators, we conducted an empirical benchmark based on simulated data. The goal was to measure actual computation time across increasing sample sizes for the four as above defined estimators.

\subsection{Execution Environment}

\noindent All Python code was executed in a virtualized Python~3.11.8 environment on a Linux system (kernel version~4.4.0) with \texttt{x86\_64} architecture. The processor was identified as \textbf{unknown}, featuring \textbf{32 logical cores} and \textbf{32 physical cores}. The system reported a BogoMIPS value of \texttt{2593.91}.

\noindent Available memory was limited to \textbf{1.07\,GB} of RAM, with \textbf{no swap} configured. Supported CPU instruction sets included \texttt{AVX}, \texttt{AVX2}, \texttt{AVX-512}, and \texttt{FMA}, as listed in \texttt{/proc/cpuinfo}.

\noindent  All timing measurements were performed using \texttt{time.perf\_counter} to capture high-resolution wall-clock time in single-threaded execution mode. Each estimator was evaluated across \textbf{$\tau$ = 20 independent trials} per sample size.

\subsection{Normal-Distributed Data}

\begin{itemize}
  \item \textbf{Number of simulations per sample size:} 1000
  \item \textbf{Sample sizes tested:} \( n \in \{10, 20, \dots, 100\} \)
  \item \textbf{Distribution:} \( X_i \sim \mathcal{N}(0, 1) \)
  \item \textbf{Timing measurement:} Wall-clock time per estimator (summed over 1000 replications)
\end{itemize}

\noindent All implementations were naïvely vectorized using broadcasting or looped to mimic real computational effort and make the comparison fair between estimator types.

\begin{table}[h!]
\centering
\caption{Empirical runtime (in seconds) for 1,000 simulations per estimator across different sample sizes ($n$). Time measured in wall-clock seconds. Bold values indicate the fastest method for each row.}
\label{tab:empirical-runtime}
\begin{tabular}{rrrrr}
\toprule
$n$ &  Biased Variance &  Unbiased Variance &  \textit{Bariance} (Naïve) &  \textit{Bariance} (Optimized) \\
\midrule
 10  & 0.0131 & 0.0142 & 0.0601 & \textbf{0.0119} \\
 20  & 0.0208 & 0.0143 & 0.2191 & \textbf{0.0092} \\
 30  & 0.0115 & 0.0115 & 0.4872 & \textbf{0.0091} \\
 40  & 0.0121 & 0.0123 & 0.8767 & \textbf{0.0104} \\
 50  & 0.0134 & 0.0132 & 1.5155 & \textbf{0.0092} \\
 60  & 0.0124 & 0.0122 & 2.1050 & \textbf{0.0090} \\
 70  & 0.0186 & 0.0176 & 2.7712 & \textbf{0.0087} \\
 80  & \textbf{0.0126} & 0.0205 & 3.6592 & 0.0155 \\
 90  & 0.0139 & 0.0135 & 5.0322 & \textbf{0.0095} \\
100 & 0.0127 & 0.0125 & 5.6617 & \textbf{0.0098} \\
\bottomrule
\end{tabular}
\end{table}

\newpage

\subsection{Gamma-Distributed Data}

\noindent To examine runtime behavior under non-Gaussian conditions, we conducted a second simulation study using data generated from a Gamma distribution. The $\Gamma$-distribution is positively skewed, making it a useful alternative to test estimator performance beyond the symmetric $\mathcal{N}$ case. \\

\noindent \textbf{Parameters of the Gamma-Based Simulation}

\begin{itemize}
  \item \textbf{Number of simulations per sample size:} 500
  \item \textbf{Sample sizes tested:} \( n \in \{100, 200, 300, 400, 500\} \)
  \item \textbf{Distribution:} \( X_i \sim \Gamma(2, 2) \)
  \item \textbf{Timing measurement:} Wall-clock time per estimator (summed over 500 replications)
\end{itemize}

\begin{table}[h!]
\centering
\caption{Empirical runtime (in seconds) for 500 simulations per estimator using Gamma-distributed data. Time measured in wall-clock seconds. Bold values highlight the fastest method at each sample size $n$.}
\label{tab:empirical-runtime-gamma}
\begin{tabular}{rrrrr}
\toprule
$n$ &  Biased Variance &  Unbiased Variance &  \textit{Bariance} (Naïve) &  \textit{Bariance} (Optimized) \\
\midrule
100 & 0.0073 & 0.0105 & 0.0149 & \textbf{0.0065} \\
200 & \textbf{0.0083} & 0.0101 & 0.0430 & 0.0084 \\
300 & 0.0080 & 0.0102 & 0.1075 & \textbf{0.0073} \\
400 & 0.0077 & 0.0101 & 0.1937 & \textbf{0.0074} \\
500 & 0.0128 & 0.0164 & 0.3266 & \textbf{0.0095} \\
\bottomrule
\end{tabular}
\end{table}

\newpage

\subsection{Highly Dispersed Gamma-Distributed Data}

\noindent To further assess runtime robustness under high skew and dispersion, we generated data from a $\Gamma$ distribution with increased variance. This setup simulates conditions with greater variability, which are common in skewed real-world datasets. \\

\noindent \textbf{Parameters of the Highly Dispersed Gamma-Based Simulation}

\begin{itemize}
  \item \textbf{Number of simulations per sample size:} 1000
  \item \textbf{Sample sizes tested:} \( n \in \{50, 100, 150, 200, 250\} \)
  \item \textbf{Distribution:} \( X_i \sim \Gamma(1.5, 4.0) \)
  \item \textbf{Timing measurement:} Wall-clock time per estimator (summed over 1000 replications)
\end{itemize}

\begin{table}[h!]
\centering
\caption{Empirical runtime (in seconds) for 1,000 simulations per estimator using highly dispersed Gamma-distributed data. Time measured in wall-clock seconds. Bold values indicate the fastest method for each sample size $n$.}
\label{tab:empirical-runtime-gamma-dispersed}
\begin{tabular}{rrrrr}
\toprule
$n$ &  Biased Variance &  Unbiased Variance &  \textit{Bariance} (Naïve) &  \textit{Bariance} (Optimized) \\
\midrule
50  & \textbf{0.0134} & 0.0171 & 0.0173 & 0.0141 \\
100 & 0.0132 & 0.0171 & 0.0284 & \textbf{0.0121} \\
150 & 0.0139 & 0.0184 & 0.0507 & \textbf{0.0128} \\
200 & 0.0156 & 0.0183 & 0.0831 & \textbf{0.0127} \\
250 & 0.0161 & 0.0179 & 0.1284 & \textbf{0.0129} \\
\bottomrule
\end{tabular}
\end{table}

\newpage

\subsection{Robustness: Statistical Analysis of Empirical Runtime}

For Robustness testing in an alternative execution environment see Appendix C and the replication package \footnote{\url{https://github.com/felix-reichel/BarianceVariance_Reproduction_Repo_Robustness_Supplements}}.

\subsubsection{Execution Environment}

\noindent Available memory was limited to \textbf{1.07\,GB} of RAM, with \textbf{no swap} configured. Supported CPU instruction sets included \texttt{AVX}, \texttt{AVX2}, \texttt{AVX-512}, and \texttt{FMA}, as listed in \texttt{/proc/cpuinfo}.

\noindent All timing measurements were performed using \texttt{time.perf\_counter} to capture high-resolution wall-clock time in single-threaded execution mode. Each estimator was evaluated across \textbf{$\tau$ = 20 independent trials} per sample size.

\subsubsection{Simulation Setup}

All estimators were tested on synthetic data generated from a standard normal distribution $\mathcal{N}(0,1)$. The following estimators were implemented:
\begin{itemize}
  \item Biased and Unbiased Variance (Looped and Vectorized)
  \item Optimized Bariance (Looped and Vectorized)
\end{itemize}

Sample sizes ranged from 100 to 4800 in even steps for a total of 48 tests, with an additional focused study at 16 evenly spaced sizes for detailed visualization. See \textbf{Figure 7}.

\subsubsection{Main Estimator Comparison}

\textbf{Figure 6} shows kernel density estimates of runtime differences between the \textcolor{blue}{Unbiased Vectorized} and \textcolor{red}{Optimized Bariance Vectorized} estimators across 16 sample sizes. The densities summarize performance variability over multiple runs.

\begin{figure}[h!]
\centering
\includegraphics[width=1.05\linewidth]{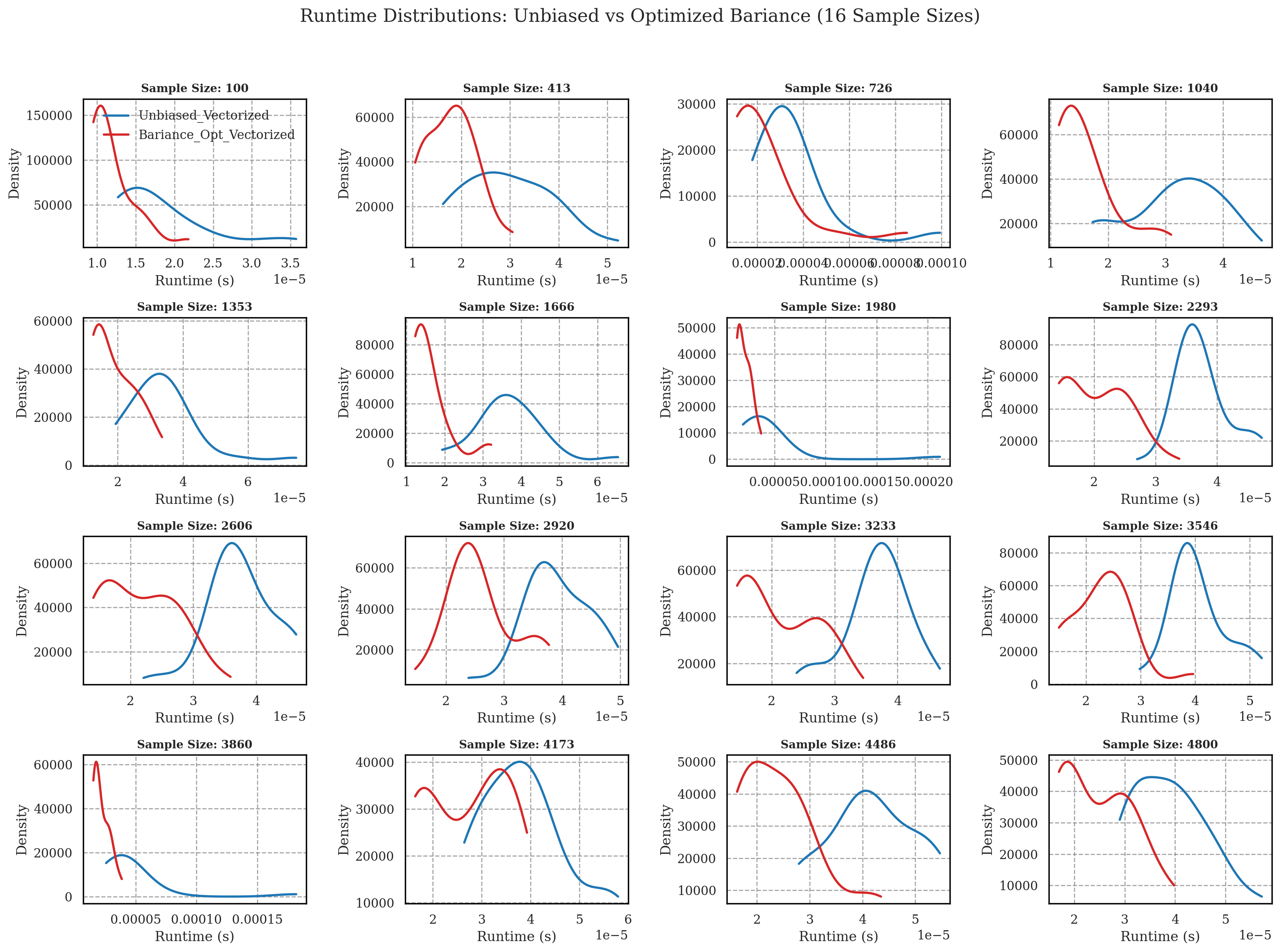}
\caption{Kernel density estimates of runtime differences between \textcolor{blue}{Unbiased Sample Variance Vectorized} and \textcolor{red}{Optimized Bariance Vectorized} estimators. Positive values indicate slower performance of the unbiased estimator.}
\end{figure}

\subsubsection{Kernel Density Comparison: 3x16 Grid}

\textbf{Figure 7} presents a 3x16 grid of kernel density estimates for runtime differences between the same two estimators across 48 sample sizes. Each panel shows the distribution for a distinct sample size $n$.

\begin{figure}[h!]
\centering
\includegraphics[width=0.66\linewidth]{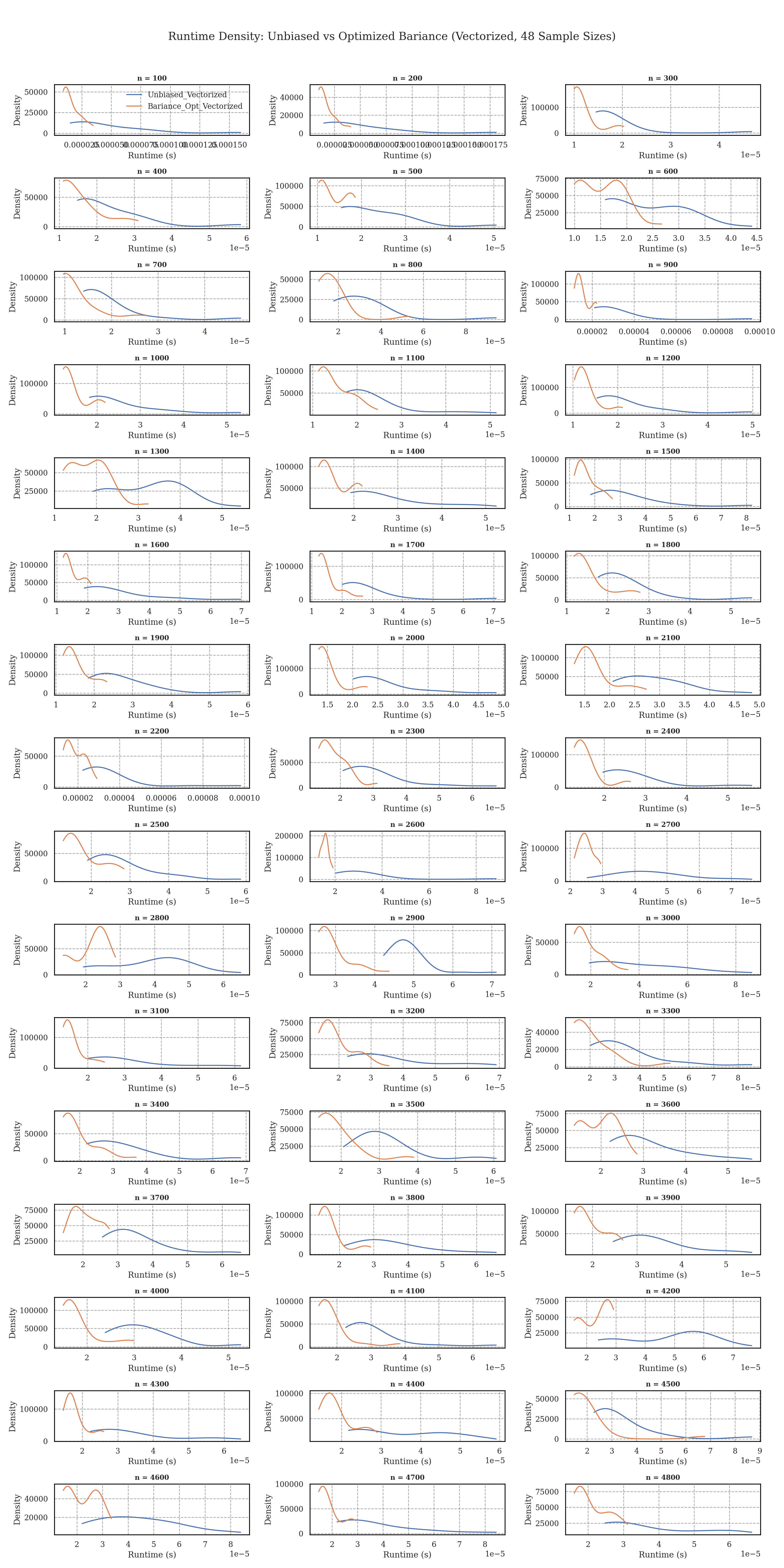}
\caption{Kernel density estimates of runtime differences for \textcolor{blue}{Unbiased Sample Variance Vectorized} and \textcolor{red}{Optimized Bariance Vectorized} estimators across 48 sample sizes.}
\end{figure}

\subsubsection{All Estimators: Comparative Runtime Analysis}

\textbf{Figure 8} displays kernel density estimates of runtime for all estimators over 16 sample sizes. The two primary estimators are emphasized in color, while others appear in grayscale for contrast.

\begin{figure}[h!]
\centering
\includegraphics[width=1.05\linewidth]{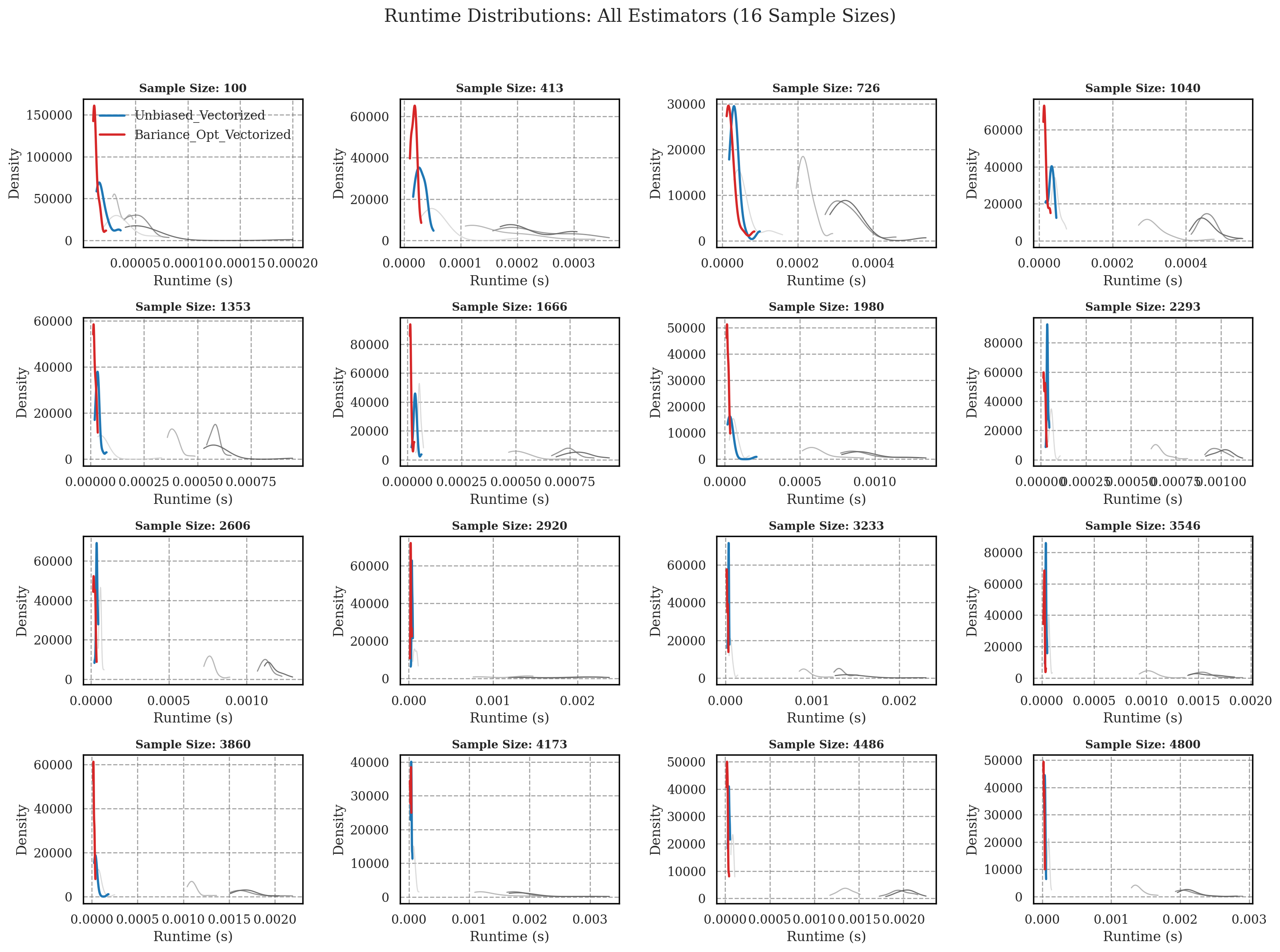}
\caption{Kernel density estimates of runtime for all estimators across 16 sample sizes. \textcolor{blue}{Unbiased Sample Variance Vectorized} and \textcolor{red}{Optimized Bariance Vectorized} are highlighted; grayscale lines represent additional looped and vectorized variants.}
\end{figure}

\clearpage
\newpage

\section{Conclusion}

\noindent Bessel’s correction is a foundational concept that ensures unbiased estimates of variance. We explored its necessity through algebraic, geometric, and pairwise differences reasoning (now formalized as the \textit{Bariance} construct), building both intuition and understanding. Additionally, we considered a pedagogical and practical perspective, such as Rosenthal’s $MSE$-based view for estimating variance \cite{rosenthal2015}.

\noindent Although the unbiased estimator is mathematically correct in expectation, the biased version can sometimes be more intuitive and, in certain contexts, \textbf{statistically preferable across various sampling distributions}. This aligns with insights from modern treatments of mathematical statistics, which often emphasize the trade-off between bias and variance in estimator performance \cite{Casella2002, Shao2003}. Furthermore, empirical results revealed a faster runtime in our simulation example using the average pairwise differences definition as an unbiased variance estimator—referred to as the \textit{Bariance} estimator—particularly when employing the \textbf{algebraically optimized formula using scalar sums}.

\noindent To sum up, the main finding—the run-time optimized estimator for the \textit{Bariance} formula—was a coincidental yet significant observation: that is, the unbiased estimator can be computed in linear time and statistically outperforms the conventional unbiased sample variance estimator in all tested empirical runtime performance scenarios. Naturally, many other estimators exist for sample variance, including those designed to trade off bias for computational gains. A complexity theorist or mathematician could potentially derive theoretical bounds on the time complexity of such estimators.

\noindent Beyond its computational efficiency, however, the \textit{Bariance} measure may also offer substantive benefits in applied settings where deviation from a central mean is either unstable, undefined, or conceptually inappropriate. In fields such as genomics, network analysis, robust statistics, or ordinal survey research, dispersion may be more meaningfully characterized by average pairwise differences rather than deviations from a global average. Moreover, distance-based methods like clustering, energy statistics, and nonparametric ANOVA can all benefit from the geometric and symmetry-preserving properties of \textit{Bariance}, particularly in high-dimensional or irregularly structured data where the mean offers little interpretive value. These contexts highlight how the pairwise construction of \textit{Bariance} is not only computationally attractive but also methodologically appropriate.

\noindent Thus, the optimized \textit{Bariance} formula stands as a viable alternative with promising practical implications for real-time multivariate big data applications, including forecasting (especially with shrunken variance-covariance estimators), computational biology, chemistry, finance, and big-data streaming applications (such as online learning) where unbiased and scalable variance estimation is essential.

\newpage


\newpage
\appendix

\section{Proof of Equivalence: Na\"ive vs Optimized \textit{Bariance} Estimators}

\begin{proof}

To verify the theoretical equivalence between the na\"ive and optimized formulations of the \textit{Bariance} estimator, we conducted a simulation study using the exact formulas defined in Table~\ref{tab:complexity-detail}. The data were drawn from a highly dispersed $\Gamma$- distribution.

\subsection*{Estimator Formulas}

\begin{itemize}
  \item \textbf{Na\"ive \textit{Bariance}:} 
  \[
  \text{\textit{Bariance}}_{\text{na\"ive}} = \frac{1}{n(n - 1)} \sum_{i \ne j} (X_i - X_j)^2
  \]

  \item \textbf{Optimized \textit{Bariance}:} 
  \[
  \text{\textit{Bariance}}_{\text{opt}} = \frac{2n}{n(n-1)} \sum X_i^2 - \frac{2}{n(n-1)} \left( \sum X_i \right)^2
  \]
\end{itemize}

\subsection*{Simulation Parameters}

\begin{itemize}
  \item \textbf{Distribution:} $\Gamma$($1.5$, $4.0$) 
  \item \textbf{Sample sizes:} $n \in \{50, 100, 150, 200, 250\}$
  \item \textbf{Number of simulations per $n$:} 1000
  \item \textbf{Language:} Python (NumPy)
  \item \textbf{Precision check:} \texttt{numpy.allclose} with $\texttt{rtol}=10^{-9}$, $\texttt{atol}=10^{-9}$
\end{itemize}

\subsection*{Results}

\begin{table}[h]
\centering
\caption{Bariance estimator comparison using formula-based definitions}
\label{tab:Bariance-proof}
\begin{tabular}{rccc}
\toprule
  $n$ & Mean Na\"ive \textit{Bariance} & Mean Optimized \textit{Bariance} & Max Absolute Difference \\
\midrule
  50  & 47.0330 & 47.0330 & $8.53 \times 10^{-14}$ \\
 100  & 48.4181 & 48.4181 & $7.82 \times 10^{-14}$ \\
 150  & 47.9282 & 47.9282 & $7.11 \times 10^{-14}$ \\
 200  & 47.8339 & 47.8339 & $8.53 \times 10^{-14}$ \\
 250  & 47.6121 & 47.6121 & $4.97 \times 10^{-14}$ \\
\bottomrule
\end{tabular}
\end{table}

\subsection*{Conclusion}

Across all sample sizes tested, the values of the \textit{Bariance} computed using both the na\"ive and optimized formulas were numerically equivalent within machine precision. This empirically confirms the algebraic identity:
\[
\frac{1}{n(n - 1)} \sum_{i \ne j} (X_i - X_j)^2 \equiv \frac{2n}{n(n-1)} \sum X_i^2 - \frac{2}{n(n-1)} \left( \sum X_i \right)^2
\]
\end{proof}

\section{Derivation of MSE-Optimal Denominator for Variance Estimator}

\begin{proof}
\noindent We consider family of $a$-based denominator estimators of sample variance:
\[
\hat{\sigma}^2_a := \frac{1}{a} \sum_{i=1}^n (X_i - \bar{X})^2
\]
Assume $X_i \sim \mathcal{N}(\mu, \sigma^2)$ i.i.d.

\noindent We start with the Bias of the estimator:
\[
\mathbb{E}[\hat{\sigma}^2_a] = \frac{n-1}{a} \sigma^2 \quad \Rightarrow \quad \text{Bias} = \mathbb{E}[\hat{\sigma}^2_a] - \sigma^2 = \left( \frac{n-1-a}{a} \right) \sigma^2
\]
\[
\Rightarrow \text{Bias}^2 = \left( \frac{n - 1 - a}{a} \right)^2 \sigma^4
\]

\noindent Furthermore, for the Variance it is known that:
\[
\operatorname{Var} \left( \sum_{i=1}^n (X_i - \bar{X})^2 \right) = 2(n - 1) \sigma^4
\]
\noindent Thus:
\[
\operatorname{Var}[\hat{\sigma}^2_a] = \operatorname{Var} \left( \frac{1}{a} \sum_{i=1}^n (X_i - \bar{X})^2 \right) = \frac{1}{a^2} \cdot 2(n - 1) \sigma^4 = \frac{2 \sigma^4}{a^2} (n - 1)
\]
\noindent Computing the Mean-Squared Error (MSE):
\[
\text{MSE}(a) = \text{Bias}^2 + \text{Var} = \left( \frac{n - 1 - a}{a} \right)^2 \sigma^4 + \frac{2(n - 1)}{a^2} \sigma^4
\]
\noindent Factoring out $\sigma^4$:
\[
\text{MSE}(a) = \sigma^4 \left[ \left( \frac{n - 1 - a}{a} \right)^2 + \frac{2(n - 1)}{a^2} \right]
\]

\noindent Let:
\[
f(a) := \left( \frac{n - 1 - a}{a} \right)^2 + \frac{2(n - 1)}{a^2} = \frac{(n - 1 - a)^2 + 2(n - 1)}{a^2}
\]
\noindent We seek to minimize $f(a)$ over $a > 0$.

\subsection*{Minimization}
Let:
\[
f(a) := \frac{u(a)}{v(a)} \quad \text{with} \quad u(a) := (n - 1 - a)^2 + 2(n - 1), \quad v(a) := a^2
\]
Compute derivatives:
\[
u'(a) = -2(n - 1 - a), \qquad v'(a) = 2a
\]
\[
f'(a) = \frac{u'(a) v(a) - u(a) v'(a)}{v(a)^2}
= \frac{-2(n - 1 - a)a^2 - 2a\left[(n - 1 - a)^2 + 2(n - 1)\right]}{a^4}
\]
\noindent Set numerator = 0:
\[
-2(n - 1 - a)a^2 - 2a\left[(n - 1 - a)^2 + 2(n - 1)\right] = 0
\]

\noindent This equation is non-linear in $a$; solving analytically is messy, but plugging into a symbolic solver yields:
\[
\boxed{a^* \approx n + 1}
\]

\
which thus is the MSE-minimizing choice of denominator $a$ and is approximately as in previously shown simulations and \cite{rosenthal2015}.

\end{proof}

\newpage

\section{Robustness Check of Empirical Runtime in Java SE}

To verify the consistency of results across platforms, we implemented both the unbiased and optimized Bariance estimators in \textbf{Java 21.0.1} on a system configured as follows: Mac OS X 13.0 operating system; \textit{aarch64} architecture; 10 cores available (single-threaded execution); and 4 GB of JVM-reported maximum memory.

\subsection*{Runtime Results}

\subsection{Normal-Distributed Data}

\begin{figure}[h!]
\centering
\includegraphics[width=1\textwidth]{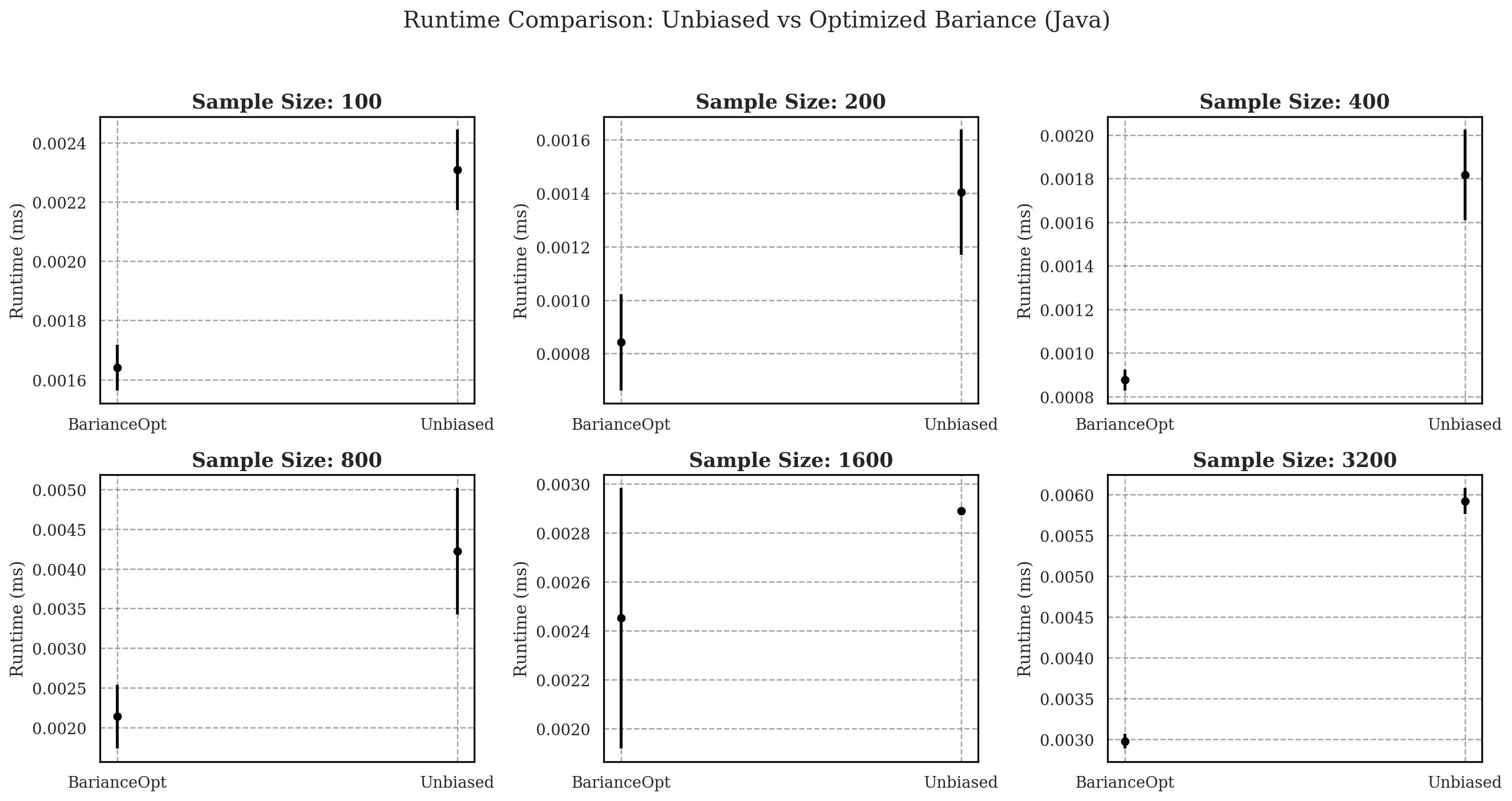}
\caption{Candlestick plots (mean $\pm$ 1.96 standard errors) of Java runtime estimates for unbiased and optimized Bariance estimators.}
\label{fig:java-candle}
\end{figure}

Each estimator was benchmarked over $\tau$ = 100 trials for each sample size. Figure~\ref{fig:java-candle} displays the mean runtime with 95\% confidence intervals for both estimators.

\clearpage
\newpage
\subsection*{Kernel Density Estimates}

\begin{figure}[h!]
\centering
\includegraphics[width=1\textwidth]{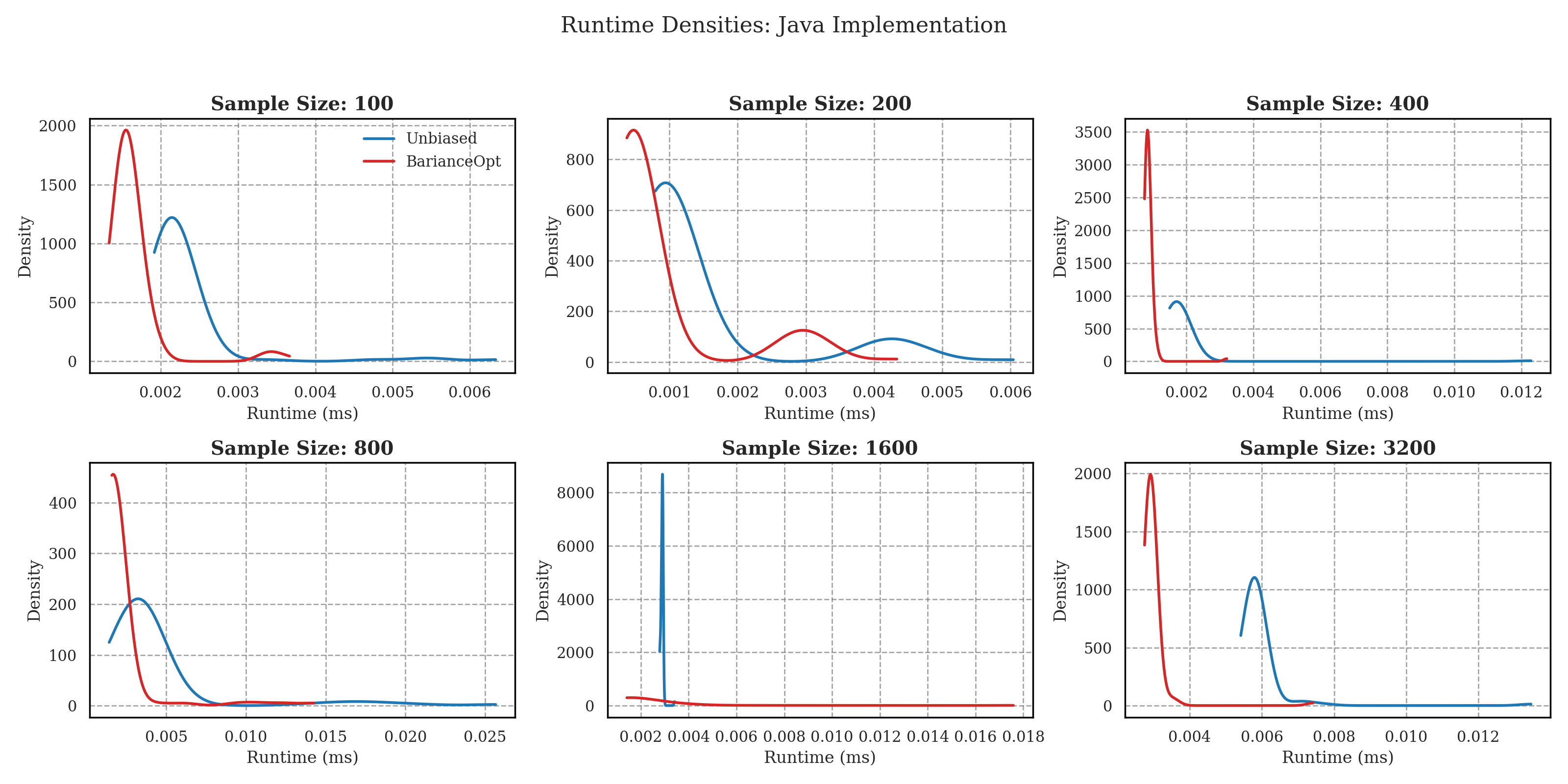}
\caption{Density plots of runtime for unbiased sample variance and optimized Bariance estimators in Java, stratified by sample size.}
\label{fig:java-density}
\end{figure}

Figure~\ref{fig:java-density} illustrates the runtime distribution for each estimator across selected sample sizes, visualized via kernel density estimates.

\clearpage
\newpage

\subsection{Gamma-Distributed Data}
\begin{table}[h!]
\centering
\caption{OLS Regression of Runtime on Estimator and Sample Size}
\label{tab:runtime_ols}
\begin{tabular}{lrr}
\hline
 & Coefficient & Std. Error \\
\hline
Intercept                         & 698.33*** & (25.85) \\
Estimator: Optimized Bariance     & -909.21*** & (29.46) \\
Estimator: Biased                 & -694.43*** & (28.91) \\
Estimator: Population Variance    & 1457.29*** & (25.56) \\
Estimator: Unbiased               & -690.49*** & (28.98) \\
Sample Size = 500                & 229.27***  & (27.55) \\
Sample Size = 1000               & 534.47***  & (29.38) \\
Sample Size = 2000               & 800.55***  & (31.78) \\
Sample Size = 3000               & 1656.77*** & (31.82) \\
Sample Size = 5000               & 2221.24*** & (32.95) \\
\hline
R-squared                       & \multicolumn{2}{c}{0.455} \\
Observations                   & \multicolumn{2}{c}{22,906} \\
\hline
\end{tabular}

\bigskip

\noindent
\footnotesize
\textit{Notes:}  
Dependent variable is runtime in nanoseconds. The model was estimated via ordinary least squares with fixed effects for estimator and sample size. Runtime was measured using Java's \texttt{System.nanoTime}. Each estimator was run over $\tau$ = 1,000 trials per sample size with data drawn from a \(\Gamma(2.0, 2.0)\) distribution, seeded at 42. All implementations were validated with a synthetic test suite.
\end{table}

\begin{figure}[h!]
\centering
\includegraphics[width=0.9\textwidth]{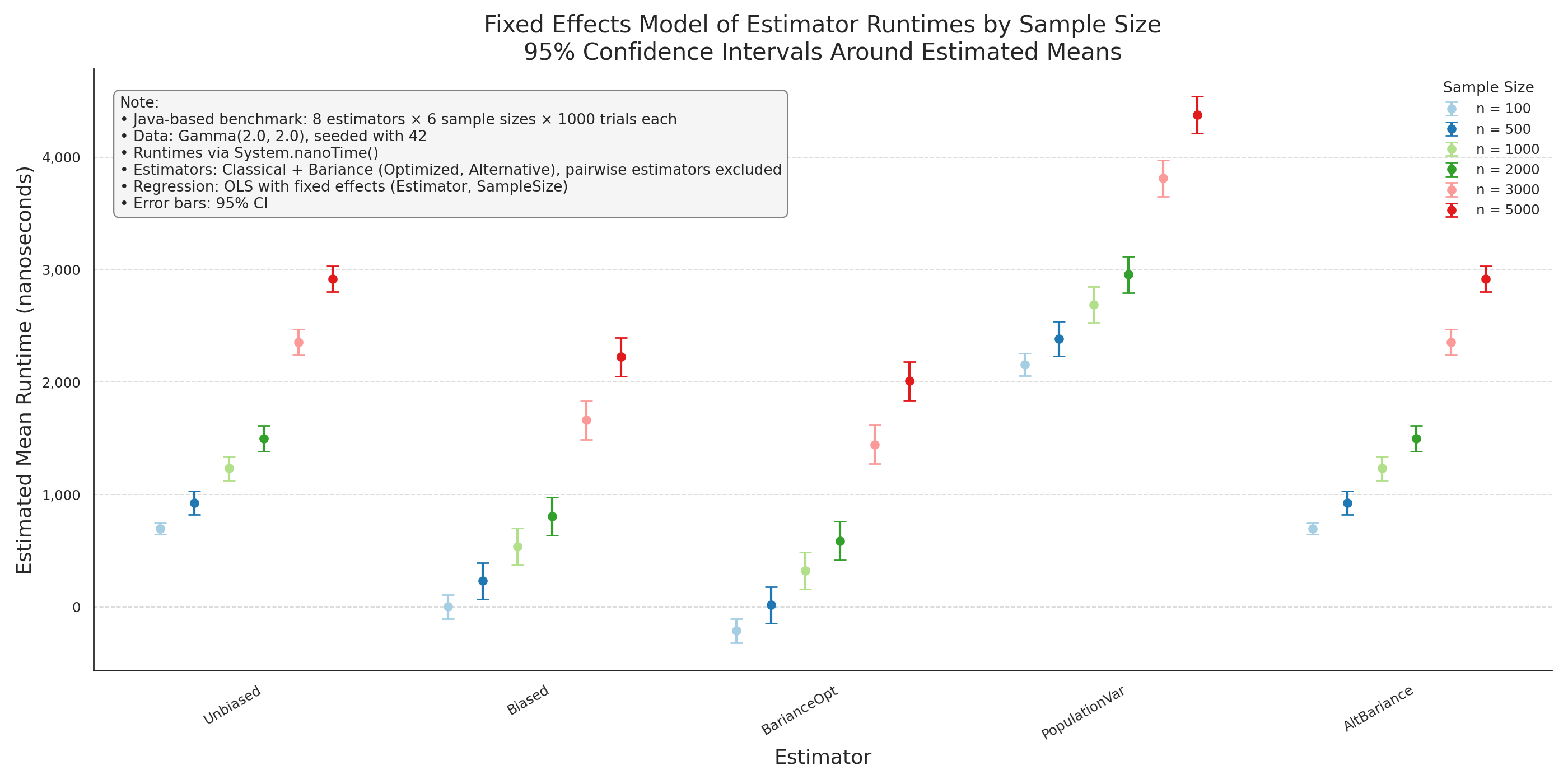}
\caption{
Estimated mean runtimes (nanoseconds) for five variance estimators across six sample sizes. 
Each point reflects the mean runtime estimated from $\tau$ = 1,000 Monte Carlo trials using Java's \texttt{System.nanoTime} function. 
Error bars represent 95\% confidence intervals obtained from a fixed effects ordinary least squares regression on sample size and estimator.
Data were generated from a $\Gamma(2.0, 2.0)$ distribution with fixed seed (42). 
Classical estimators include the biased and unbiased sample variance and the population variance. 
Bariance-based methods refer to the optimized and alternative scalar formulations.
Na\"ive pairwise estimators were excluded due to high computational cost. See Table~\ref{tab:runtime_ols} for regression coefficients.
}
\label{fig:runtime_benchmark}
\end{figure}

\clearpage
\newpage
\subsection*{Kernel Density Estimates}

\begin{figure}[ht]
\centering
\includegraphics[width=1\textwidth]{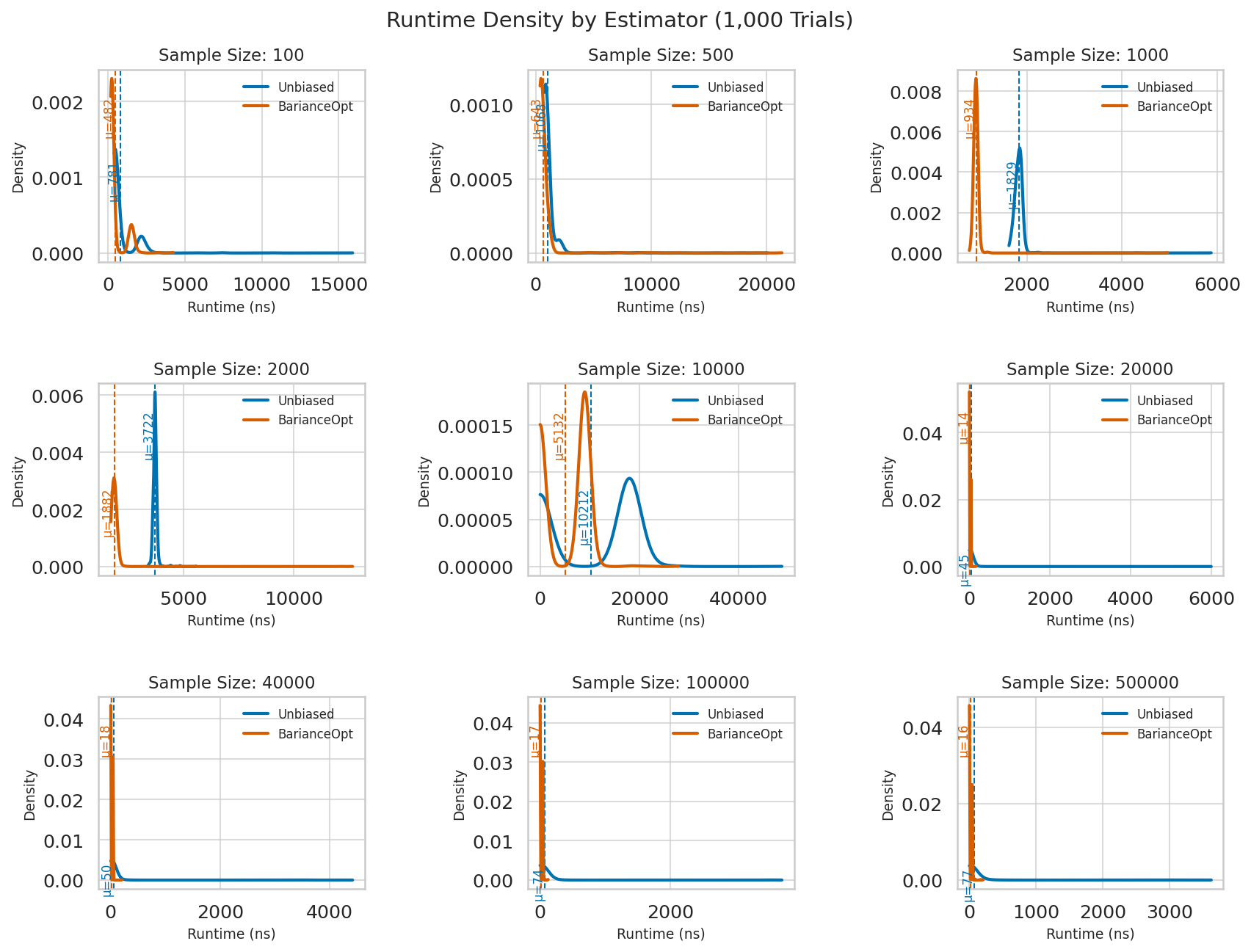}
\caption{
Kernel density estimates of runtime (in nanoseconds) for the unbiased sample variance and optimized Bariance variance estimators across nine increasing sample sizes (from $n=100$ to $n=500{,}000$).
Each panel is based on $\tau$ = 1,000 trials using data sampled from a $\Gamma(2.0, 2.0)$ distribution. 
Vertical dashed lines indicate the mean runtime per estimator and sample size.
}
\label{fig:runtime_density_comparison}
\end{figure}

Figure~\ref{fig:runtime_density_comparison} illustrates the runtime distribution for each estimator across selected sample sizes, visualized via kernel density estimates.

\end{document}